\documentclass[aps,preprint,a4paper]{revtex4-1}

\usepackage[utf8]{inputenc}
\usepackage{amsmath}
\usepackage{amsfonts}
\usepackage{amssymb}
\usepackage{epsfig}

\usepackage{relsize}
\usepackage{dsfont}
\usepackage{overpic}
\usepackage{caption}
\usepackage{subcaption}
\usepackage{moredefs}
\usepackage{color}
\usepackage{dsfont}
 
\newcommand{\eps}{\varepsilon}

\newcommand{\dx}{\, \mathrm{d}\mathbf{x}}

\newcommand{\mb}[1]{\mathbf{#1}}

\begin{document}

\title{Collective migration under hydrodynamic interactions - a computational approach}

\author{W. Marth}
\affiliation{Institut f\"ur Wissenschaftliches Rechnen, TU Dresden, 01062 Dresden, Germany 
({\tt wieland.marth@tu-dresden.de})}
\author{A. Voigt}
\affiliation{Institut f\"ur Wissenschaftliches Rechnen, TU Dresden, 01062 Dresden, Germany; Dresden Center for Computational Materials Science (DCMS), TU Dresden, 01062 Dresden, Germany; Center for Systems Biology Dresden (CSBD), Pfotenhauerstr. 108, 01307 Dresden, Germany ({\tt axel.voigt@tu-dresden.de})}

\begin{abstract}
Substrate-based cell motility is essential for fundamental biological processes, such as tissue growth, wound healing and immune response. Even if a comprehensive understanding of this motility mode remains elusive, progress has been achieved in its modeling using a whole cell physical model. The model takes into account the main mechanisms of cell motility - actin polymerization, substrate mediated adhesion and actin-myosin dynamics and combines it with steric cell-cell and hydrodynamic interactions. The model predicts the onset of collective cell migration, which emerges spontaneously as a result of inelastic collisions of neighboring cells. Each cell here modeled as an active polar gel, is accomplished with two vortices if it moves. Open collision of two cells the two vortices which come close to each other annihilate. This leads to a rotation of the cells and together with the deformation and the reorientation of the actin filaments in each cell induces alignment of these cells and leads to persistent translational collective migration. The effect for low Reynolds numbers is as strong as in the non-hydrodynamic model, but it decreases with increasing Reynolds number. 
\end{abstract}

\maketitle

\section{Introduction}

Substrate-based cell motility is a well studied process for eukaryotic cells, such as keratocytes, fibroblasts and neutrophils. It plays a fundamental role
in tissue growth, wound healing and immune response. The main processes involved in this cell motion are: (i) the generation of a propulsive force by 
actin polymerization, which act against the cell's membrane, (ii) the formation of adhesive contact to the substrate, transforming this force to the substrate 
to move forward and (iii) a contractile action of actin-myosin complexes determining the cell polarity and being responsible for retraction of the cell's rear, 
see e.g. \cite{Abercrombie_PRSL_1980,Anamthakrishnanetal_IJBS_2007} for a review on the forces involved in cell movement. Several experimental studies for fish keratocyte, e.g. \cite{Fournieretal_JCB_2010,Barnhartetal_PLoSBiol_2011,Lieberetal_CB_2013},
indicate a self-organization process behind the motility mechanism, which has been adapted in various theoretical approaches \cite{Tjhungetal_PNAS_2012,
Ziebertetal_JRSInterface_2012,Giomietal_PRL_2014,Whitfieldetal_EPJE_2014,Marthetal_JRSInterface_2015}. They all apply an active 
polar gel theory \cite{Kruseetal_PRL_2000,Kruseetal_PRL_2004,Kruseetal_EPJE_2005}. If considered in a  confinement, a splayed polarization of the 
actin filaments can occur, which models the contractile stress due to the interaction of myosin and actin. If combined with the treadmilling 
process of polymerization and depolymerization of actin filaments, as e.g. considered in \cite{Shaoetal_PRL_2010,Shaoetal_PNAS_2012,Marthetal_JMB_2014} 
and an effective treatment of the adhesive contact, a whole-cell physical model for moving cells can be constructed 
\cite{Tjhungetal_NC_2015,Marthetal_JRSInterface_2015}. Such models have been established for 
single cells and used to analyze motility of various cell types \cite{Loeberetal_SM_2014,Marthetal_JRSInterface_2015}. The results strongly support the 
physical view on cellular motility, which exploits autonomous physical mechanisms whose operation does not need continuous regulatory effort. Recently 
such models have also been considered for collective migration \cite{Loeberetal_SR_2015}. Here each cell is considered as an active polar gel and interactions
between the cells are specified. The model predicts that collective migration emerges spontaneously as a result of inelastic collisions between neighboring cells. 
These collisions lead to mutual alignment of the cells velocities and to the formation of coherently-moving multi-cellular clusters. These results essentially confirm 
simpler agent-based modeling approaches of Vicsek-type \cite{Vicseketal_PRL_1995} with inelastic behaviour in the interaction rules \cite{Grossmanetal_NJP_2008}, 
recent mesoscopic simulations based on active phase field crystal models \cite{Alaimoetal_arxiv_2016} and continuum approaches, which only consider the 
emerging macroscopic behaviour \cite{Wittkowskietal_NC_2014,Specketal_PRL_2014} using Cahn-Hilliard type models. All these approaches for collective migration
neglect hydrodynamic interactions, which are of widespread importance for cells. The effect of these interaction on collective migration is controversially discussed.
In the related problem of motility induced phase separation \cite{Catesetal_ARCMP_2015}, where clustering results from to a reduction of the propulsion speed due to cell
collisions in environments with high local density, \cite{Matas-Navarroetal_PRE_2014,Matas-Navarroetal_SM_2015}, a suppression of cluster formation is observed if 
hydrodynamic interaction is taken into account, while the hydrodynamic active Cahn-Hilliard model in \cite{Tiribocchietal_PRL_2015} leads to arrested phase 
separation.

We here consider the hydrodynamic active polar gel model, which was used in \cite{Marthetal_JRSInterface_2015} for a single cell, for multiple cells. Each cell is thereby
described by a phase field variable, which defines the confinement of the field variables of the active polar gel model for each cell. The
interaction between the cells only considers steric interactions. Short range repulsion is realized by a Gaussian potential using the phase field variables \cite{Marthetal_JFM_2016}.
Using a multi-mesh approach \cite{Lingetal_CMAM_2016}, which allows for an efficient numerical treatment by considering differently refined meshes for each variable, allows to
significantly reduce the computational cost and to consider numbers of cells, which are sufficient for collective migration. 

The paper is organized as follows: In Section~\ref{sec:mathmodelsactive} we introduce the mathematical model
and compare it with the non-hydrodynamic model in \cite{Loeberetal_SR_2015}. We further discuss numerical aspects. In Section~\ref{sec:simulations}, we first perform several 
computations for binary collisions before the onset of collective migration is studies for larger systems. The simulations do not indicate a suppression of collective motion if hydrodynamic interactions are considered.

\section{Mathematical model for substrate-based cell motility}\label{sec:mathmodelsactive}

The mathematical model is based on physical phenomena and results from energy minimization, conservation laws and active components, taking into account the filament network, the cell membrane, cell-cell and cell-substrate interactions, as well as fluid properties.

\subsection{Energy}
Following \cite{Tjhungetal_PNAS_2012,Marthetal_JRSInterface_2015} we consider the free energy of a single cell $i$
\begin{equation}
 \mathcal{E}_{cell}(\mathbf{P}_i, \phi_i)=\mathcal{E}_P(\mb P_i,\phi_i) + \mathcal{E}_S(\phi_i)
\end{equation}
which consists of the energy of the filament network $\mathcal{E}_P(\mb P_i,\phi_i)$, described by an orientation field $\mathbf{P}_i$, which is the mesoscopic average of the actin filaments and the surface energy $\mathcal{E}_S(\phi_i)$ of the cell membrane $\Gamma_i(t)$. Each cell is described by a phase field variable $\phi_i$, defined as
$\phi_i(t,\mathbf{x}):=\tanh (r_i(t,\mathbf{x})/(\sqrt{2}\epsilon))$, where $\epsilon$ characterizes the thickness of the diffuse interface and $r_i(t,\mathbf{x})$ denotes the signed-distance function between $\mathbf{x} \in \Omega$, in the considered case a bounded domain in $I\!\!R^2$, and its nearest point on $\Gamma_i(t)$. Depending on $r_i$, we label cell $i$ with $\phi_i\approx 1$ and the outside with $\phi_i\approx -1$. The cell membrane $\Gamma_i(t)$ is then implicitly defined by the zero level set of $\phi_i$. In  \cite{Tjhungetal_PNAS_2012} the cell has been considered as a droplet for which the surface energy reads 
\begin{align}
 \mathcal{E}_{S,CH}(\phi_i) & =\frac{3\sigma_i}{2\sqrt{2}}\int_\Omega \frac{\eps}{2}|\nabla \phi_i|^2+\frac{1}{\eps} W(\phi_i)\dx
\end{align}
where $W(\phi_i)=\frac{1}{4}(\phi_i^2-1)^2$ denotes the double-well potential and $\sigma_i$ is the membrane tension. In \cite{Marthetal_JRSInterface_2015} also a bending energy of the cell membrane was taken into account using the Helfrich energy in a phase-field approximation \cite{Duetal_NONL_2005,Hausseretal_IJBMBS_2013}
\begin{align}
 \mathcal{E}_{S,W}(\phi_i) & = \frac{3b_{N,i}}{4\sqrt{2}} \int_\Omega\frac{1}{2\eps}\left(\eps \Delta\phi_i-\frac{1}{\eps}W_{0}^\prime(\phi_i)\right)^2\dx
\end{align}
with $b_{N,i}$ denoting the bending rigidity and $W_{0,i}^\prime(\phi_i)=(\phi_i^2-1)(\phi_i+\sqrt{2}H_{0,i}\eps)$ the derivative of the double-well potential with the spontaneous curvature $H_{0,i}$. The surface energy thus results as a combination of both energies
\begin{align}
 \mathcal{E}_S(\phi_i)=\mathcal{E}_{S,CH}(\phi_i)+\mathcal{E}_{S,W}(\phi_i).
\end{align}
In the following we will consider $\sigma_i = \sigma$, $b_{N,i} = b_N$ and $H_{0,i} = H_0$ for simplicity.
The energy of the filament network of cell $i$ is given by 
\begin{align}
 \mathcal{E}_P(\mb P_i,\phi_i)=\int_\Omega \frac{k_i}{2}(\nabla\mathbf{P}_i)^2 + \frac{c_{0,i}}{4} |\mathbf{P}_i|^2 (- 2 \phi_i + |\mathbf{P}_i|^2) +\beta_{0,i}\mathbf{P}_i\cdot\nabla\phi\dx.\label{eq:energ}
\end{align}
The gradient term with the positive Frank constant $k_i$ is a simplification of a general distortion energy formulation from the theory of liquid crystals, with the assumption of the same value of the stiffness associated with splay and bend deformations, see e.g. \cite{deGennesetal_1993}.  Linking $\phi_i$ to the second term allows restricting $\mathbf{P}_i$ to the cytoplasm: If $\phi_i<0$ the minimum is obtained for $|\mathbf{P}_i| = 0$ and thus the term does not contribute to the energy, and for $\phi_i>0$ the term forms a double-well with two minima with $|\mathbf{P}|=1$ and the form specified by the parameter $c_{0,i}$.  The last term in eq. (\ref{eq:energ}) guarantees for $\beta_{0,i}>0$ that $\mathbf{P}_i$ points outwards in normal direction to the cell boundary. This is required to account for the effect of polymerization of actin filaments \cite{Ziebertetal_PONE_2013}. We will again only consider the case $k_i = k$, $c_{0,i} = c_0$ and $\beta_{0,i} = \beta_0$. 

\begin{figure}[h]
\begin{center}
\includegraphics[width=0.5\textwidth]{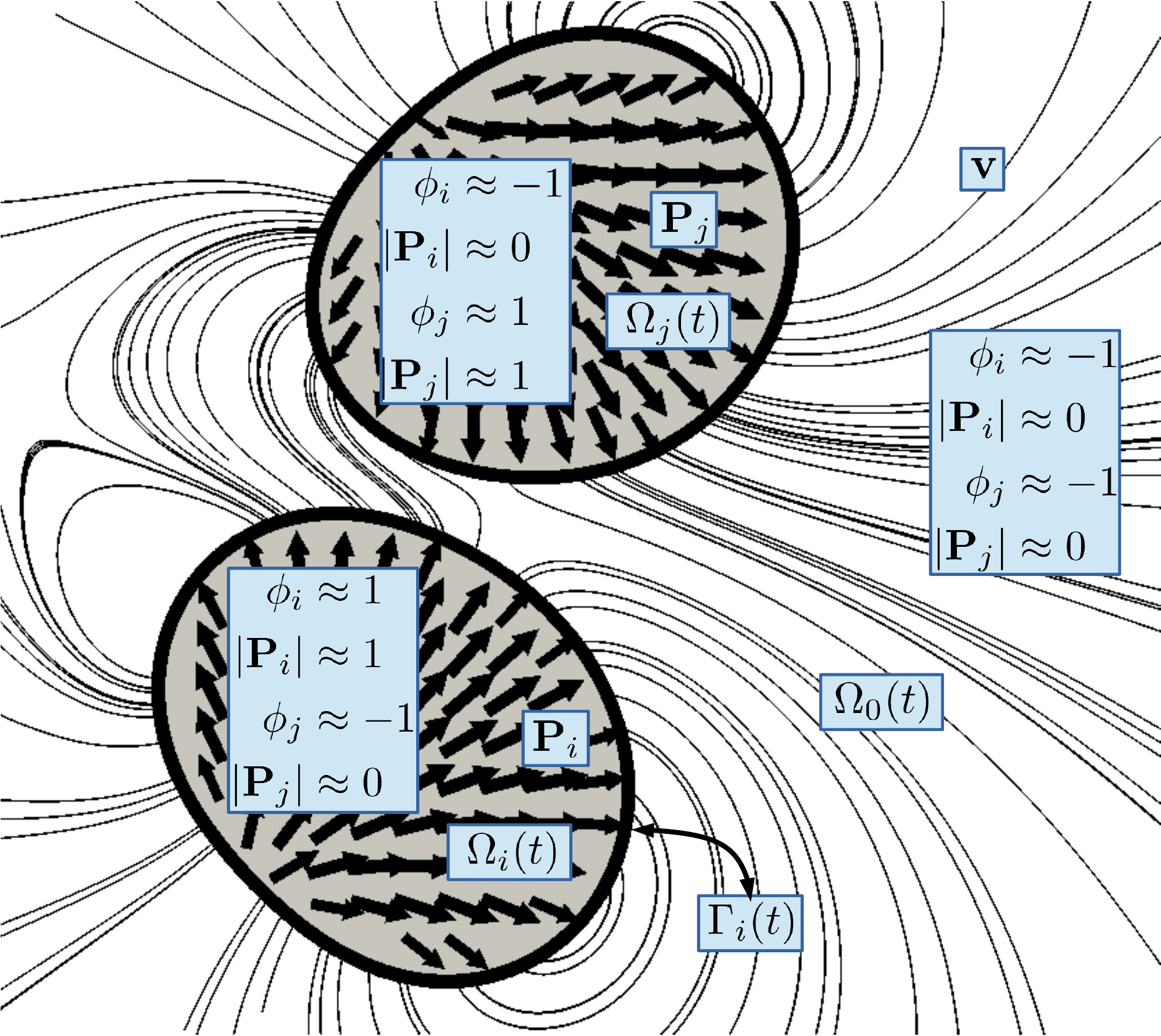}
\end{center}
\caption{Schematic description for two moving cells. Shown are the splayed orientation field $\mathbf{P}_{i,j}$, as well as the streamlines of the velocity profile $\mathbf{v}$ and the phase-fields $\phi_{i,j}$ with the cell membranes $\Gamma_{i,j}(t)$ corresponding to the zero-level sets of $\phi_{i,j}$. (Online version in colour.)}
\label{fig1}
\end{figure}

The overall energy for $N$ cells and their interaction in a fluid environment is given by
\begin{equation*}
 \mathcal{E}(\mathbf{P}_1, \ldots, \mathbf{P}_N,\phi_1, \ldots, \phi_N, \mathbf{v})=\sum_{i=1}^N \mathcal{E}_{cell}(\mb P_i,\phi_i) + \sum_{i=1}^N \mathcal{E}_{i,int}(\phi_1,\ldots,\phi_N) + \mathcal{E}_{kin}(\mb v) 
\end{equation*}
with the kinetic energy $\mathcal{E}_{kin}$ and the velocity $\mathbf{v}$. For the sake of simplicity, we consider in the derivation equal density $\rho$ and viscosity $\eta$ for $\Omega_{cell}(t)=\cup_{i=1}^N \Omega_i(t)$ and the fluid outside $\Omega_{0}(t)$, which is considered as an isotropic Newtonian fluid, so that 
\begin{align}
\mathcal{E}_{kin}(\mathbf{v})=\frac{\rho}{2}\int_\Omega \mathbf{v}^2 \dx
\end{align} 
with $\Omega = \Omega_{0}(t) \cup \Gamma(t) \cup \Omega_{cell}(t)$ and $\Gamma(t) = \cup_{i=1}^N \Gamma_i(t)$. We further introduce the phase field $\phi_{cell}=\max(\phi_1,\ldots,\phi_N)$ containing all cells. Fig. \ref{fig1} provides a schematic description for two cells.

The cell-cell interaction energy ${\cal{E}}_{i,int}$ requirers a coupling of all surrounding phase fields $\phi_1, \ldots,\phi_{i-1},\phi_{i+1},\ldots, \phi_N$ with $\phi_i$. We here consider only steric interactions and model a short range repulsion by a Gaussian potential. Following \cite{Marthetal_JFM_2016} we use the definition of $\phi_j(t,\mathbf{x}):=\tanh (r_j(t,\mathbf{x})/(\sqrt{2}\epsilon))$ to compute the signed distance function $r_j$, which is used to link cell i and cell j. Within the diffuse interface region we obtain
\begin{equation}
r_{j}= - \frac{\epsilon}{\sqrt{2}} \ln \frac{1+\phi_j}{1-\phi_j}  \quad \forall \mathbf{x}: |\phi_j(\mathbf{x})| < 1
\end{equation}
and thus can write the Gaussian interaction potential within the phase-field description as
\begin{equation}
\mathcal{E}_{i,int} (\phi_1, \ldots, \phi_N) = \int_{\Omega} B(\phi_i) \sum_{j=1 \atop j \ne i}^N \alpha_{ij} w_{j} d \Omega
\end{equation}
with $B(\phi_i) = \frac{1}{\epsilon} (\phi_i^2 - 1)^2$ being nonzero only within the diffuse interface around $\Gamma_i$, the interaction function
\begin{equation}
w_{j} =
\begin{cases}
  \exp \left((- \frac{1}{2} (\ln \frac{1+\phi_j}{1-\phi_j})^2\right),  & \text{if } |\phi_j(\mathbf{x})|<1\\
  0 & \text{otherwise}
\end{cases}
\end{equation}
and $\alpha_{ij} > 0$ the strength of the repulsive interaction between cell $i$ and cell $j$ with respect to the evolution of cell $i$. Here, we consider a constant repulsive interaction strength, hence $\alpha_{ij} = \alpha$. The approach circumvents any non-local terms which are typically required for cell-cell interactions and has been analyzed in detail in \cite{Marthetal_JFM_2016}.

\subsection{Non-dimensional form}
Before we introduce the governing equations, we consider the energies in a non-dimensional form. We consider the characteristic values for space $\mathbf{x}=L \mathbf{\hat x}$, velocity $\mathbf{v} = V \mathbf{\hat v}$ and energy $E=\eta V L^2 \hat E$, with characteristic length $L$, characteristic velocity $V$ and fluid viscosity $\eta$. This yields a time scale $t=L/V \hat t$ and a pressure $p=\eta V/L  \hat  p$. We further define the constants $c = c_{0} L^2 / k$ and $\beta= \beta_{0} L/k$ and the dimensionless quantities: 
\begin{align*}
\text{Re}\;= \frac{\rho U L}{\eta}, \quad \text{Ca}=\frac{ 2\sqrt{2}}{3}\frac{\eta U}{\sigma}, \quad \text{Be}=\frac{4\sqrt{2}}{3}\frac{\eta U L^2}{b_{N}}, \quad \text{Pa}=\frac{\eta U L }{k}, \quad \text{In} = \frac{4 \sqrt{2}}{3} \frac{\eta U}\alpha
\end{align*}
which are Reynolds, Capillary, Bending capillary, Polarity and Interaction number, respectively. Dropping the ${\hat \cdot}$ notation we obtain the energies in a non-dimensional form
\begin{align*}
\mathcal{E}_{P}(\mb P_i,\phi_i)&=\frac{1}{\text{Pa}}\int_\Omega \frac{1}{2}(\nabla\mathbf{P}_i)^2+\frac{c}{4}|\mathbf{P}_i|^2 (-2\phi_i +|\mathbf{P}_i|^2)+\beta\mathbf{P}_i\cdot\nabla\phi_i\dx \\
\mathcal{E}_{S,CH}(\phi_i) &=\frac{1}{\text{Ca}}\int_\Omega \frac{\eps}{2}|\nabla \phi_i|^2+\frac{1}{\eps} W(\phi_i)\dx\\
\mathcal{E}_{S,W}(\phi_i) &=\frac{1}{\text{Be}} \int_\Omega\frac{1}{2\eps}\left(\eps \Delta\phi_i-\frac{1}{\eps}W_{0}^\prime(\phi_i)\right)^2 \!\!\! \dx\\
\mathcal{E}_{kin}(\mb v)&=\frac{\text{Re}}{2}\int_\Omega \mathbf{v}^2 \dx\\
\mathcal{E}_{i,int}(\phi_1, \ldots, \phi_N)&= \frac{1}{\text{In}} \int_{\Omega} B(\phi_i) \sum_{j=1\atop j\ne i}^N w_{j} \dx,
\end{align*}
and again $\mathcal{E}_{S}(\phi_i) = \mathcal{E}_{S,CH}(\phi_i) + \mathcal{E}_{S,W}(\phi_i)$, $\mathcal{E}_{cell}(\mathbf{P}_i, \phi_i)=\mathcal{E}_P(\mb P_i,\phi_i) + \mathcal{E}_S(\phi_i)$ and $\mathcal{E}(\mathbf{P}_1,\ldots,\mathbf{P}_N,$ $\phi_1, \ldots, \phi_N, \mathbf{v})=\sum_{i=1}^N \mathcal{E}_{cell}(\mb P_i,\phi_i) + \sum_{i=1}^N \mathcal{E}_{i,int}(\phi_1,\ldots,\phi_N) + \mathcal{E}_{kin}(\mb v)$. 

\subsection{Governing equations}
The hydrodynamic model is an extension of the model in \cite{Tjhungetal_PNAS_2012,Marthetal_JRSInterface_2015}. The governing equations look similar, but now have to be considered for each cell with the additional contribution from the interaction terms. We denote the variational derivative or chemical potential of the orientation fields and phase fields by $\mathbf{P}^\natural_i = \delta \mathcal{E} / \delta \mathbf{P}_i$ and $\phi^\natural_i = \delta \mathcal{E} / \delta \phi_i$.

The evolution equations for the phase field variables $\phi_i$ are regularized advection equations with the advected velocity given by the fluid velocity $\mathbf{v}$. The introduced diffusion
term is scaled with a small mobility coefficient $\gamma>0$. The equations read
\begin{align}
\partial_{t} \phi_i + \mathbf{v}\cdot \nabla \phi_i  & = \gamma \Delta \phi^\natural_{i}, \quad i = 1, \ldots, N
\end{align}
and are coupled with each other through the fluid velocity $\mathbf{v}$ and the interaction terms, which are contained in the chemical potentials $\phi^\natural_{i}$, which read
\begin{align*}
\phi^\natural_{i} & =\frac{1}{\text{Be}}\left(\Delta\mu_i-\frac{1}{\eps^2}W_{0}^{\prime\prime}(\phi_i)\mu_i\right)+ \frac{1}{\text{Ca}}\left(-\eps\Delta\phi_i+\frac{1}{\eps}W^{\prime}(\phi_i)\right)\nonumber\\
& \quad +\frac{1}{\text{Pa}}\left(- \frac{c}{2}|\mb P_i|^2-\beta\nabla\cdot \mb P_i\right)+\frac{1}{\text{In}}\left(B^\prime(\phi_i)\sum_{j=1\atop j\ne i}^N w_j +w_i^\prime \sum_{j=1\atop j\ne i}^N B(\phi_j)\right) \label{eq:phimodel1b2} \\
\mu_i &= \epsilon \Delta \phi_i - \frac{1}{\epsilon} W^\prime(\phi_i)
\end{align*}
for $i = 1, \ldots, N$.

The orientation field equations for each $\mb P_i$ are the same as for the single cell case and read
\begin{align}
\partial_{t}\mathbf{P}_i + (\mathbf{v} \cdot  \nabla)\mathbf{P}_i +  \mathbf{\Omega} \cdot \mathbf{P}_i & =  \xi \mathbf{  D} \cdot \mathbf{P}_i - \frac{1}{\kappa}\mathbf{P}_i^\natural, \quad i = 1, \ldots, N
\end{align}
where the left hand side is the co-moving and co-rotational derivative where the vorticity tensor defined as $\mathbf{\Omega}=\frac{1}{2}(\nabla \mathbf{v}^\top-\nabla \mathbf{v})$ takes rotational effects from the flow field into account. The first term on the right hand side describes the alignment of $\mathbf{P}_i$ with the flow field, with the deformation tensor $\mathbf{D}=\frac{1}{2}(\nabla \mathbf{v}+\nabla \mathbf{v}^\top)$. $\xi$ and $\kappa$ are non-dimensional material parameters The evolution equations are defined in $\Omega$, but due to the coupling with $\phi_i$ we have $|\mathbf{P}_i| \approx 0$ outside of cell $i$. The non-dimensional chemical potentials read
\begin{align*}
\mathbf{P}_i^\natural & = \frac{1}{\text{Pa}}\left(-c \phi_i \mathbf{P}_i + c \mathbf{P}_i^2 \mathbf{P}_i -  \Delta \mathbf{P}_i+\beta \nabla  \phi_i\right), \quad i = 1,\ldots, N.
\end{align*}

The flow field $\mathbf{v}$ and pressure $p$ are defined through the incompressible Navier-Stokes equations, which read
\begin{align}
\text{Re}(\partial_t \mathbf{v} + (\mathbf{v}\cdot \nabla)\mathbf{v}) + \nabla p & = -\theta\mathbf{v} + \nabla\cdot \boldsymbol{\sigma} + \mathbf{F} \\
\nabla\cdot\mathbf{v} &  = 0,
\end{align}
with friction coefficient $\theta$, modeling substrate adhesion, hydrodynamic stress tensor $\boldsymbol{\sigma} = \boldsymbol{\sigma}_{viscous} + \boldsymbol{\sigma}_{active} + \boldsymbol{\sigma}_{dist} + \boldsymbol{\sigma}_{ericksen}$, consisting of passive and active components, and a forcing term $\mathbf{F}_{poly}$. The viscous stress is
\begin{align}
\boldsymbol{\sigma}_{viscous} = \eta(\phi_{\text{cell}}) \mathbf{D},
\end{align}
with $\phi_{\text{cell}} = \sum_{i = 1}^N (\phi_i +1) - 1$ and $\eta(\phi_\text{cell}) = 1$ if the outer fluid and the cells have the same viscosity and a quotient if they differ. The active stress due to actin-myosin complexes is
\begin{align}
\boldsymbol{\sigma}_{active} = \sum_{i=1}^N \frac{1}{Fa} \mathbf{P}_i \otimes \mathbf{P}_i,
\end{align}
with the active force number $Fa = \eta V / \xi L$ and $\xi > 0$. The stress coming from the distortions of the filaments, reads
\begin{align}
\boldsymbol{\sigma}_{dist} = \sum_{i=1}^N \left( \frac{1}{2}(\mathbf{P}_i^\natural \otimes\mathbf{P}_i -\mathbf{P}_i\otimes \mathbf{P}_i^\natural) + \frac{\xi}{2}(\mathbf{P}_i^\natural\otimes \mathbf{P}_i + \mathbf{P}_i \otimes\mathbf{P}_i^\natural)\right),
\end{align}
and for the Ericksen stress we consider the divergence to be defined through
\begin{align}
\nabla \cdot \boldsymbol{\sigma}_{ericksen} = \sum_{i=1}^N \phi^\natural_i \nabla \phi_i + \sum_{i=1}^N \nabla \mathbf{P}^T_i \cdot \mathbf{P}^\natural_i.
\end{align}
The forcing term accounts for actin polymerization and reads $\mathbf{F}_{poly} = \sum_{i=1}^N v_{0,i} \mathbf{P}_i$, with the non-dimensional self-propulsion velocity $v_{0,i}$. We again only consider the case $v_{0,i} = v_0$.

If we set $N = 1$, we obtain the system considered in \cite{Marthetal_JRSInterface_2015} with two additional terms in the Navier-Stokes equations. The first is the friction term $\theta \mathbf{v}$, which has not been considered as the focus in \cite{Marthetal_JRSInterface_2015} is on motility in environments without local adhesion, and the second is the forcing term $\mathbf{F}_{poly}$, as actin polymerization is not taken into account in \cite{Marthetal_JRSInterface_2015}. However, both terms had already been considered in \cite{Tjhungetal_PNAS_2012}.

\subsection{Non-hydrodynamic model}
For comparison we consider also a non-hydrodynamic model. As all stress and forcing terms has been considered in the Navier-Stokes equations, we cannot simply neglect the hydrodynamic interactions. Instead we consider 
\begin{align}
\partial_{t} \phi_i +  v_{0}\mb P_i \cdot \nabla \phi_i  & = \gamma \Delta \phi^\natural_{i}, \quad i = 1, \ldots, N \\
\partial_{t}\mathbf{P}_i + (v_{0} \mathbf{  P}_i \cdot  \nabla)\mathbf{P}_i&  =  - \frac{1}{\kappa}\mathbf{P}_i^\natural, \quad i = 1, \ldots, N,
\end{align}
with the advections only due to the self-propelled velocity $v_{0}$. The chemical potentials $\phi^\natural_{i}$ and $\mathbf{P}^\natural_i$ are defined as before. This model can be related to the model used for collective migration in \cite{Loeberetal_SR_2015}. However, several differences should be point out. We here neglect the treatment of adhesion bonds and the viscoelastic properties of the substrate. Furthermore the cell-cell interaction is considered differently. We do only consider steric interactions and no cell-cell adhesion. However, the strongest difference is the treatment of the orientation fields $\mathbf{P}_i$. In \cite{Loeberetal_SR_2015} only one variable is used for all cells. As the equation contains diffusion/elasticity of the orientation field this induces an unphysical coupling of the actin filaments over cell boundaries. 

\subsection{Numerical approach and implementation}\label{sec:numerics}
The system of partial differential equations is discretized using the parallel adaptive finite element toolbox AMDiS \cite{Veyetal_CVS_2007,Witkowskietal_ACM_2015}. We use a semi-implicit time discretization and an operator splitting approach that allows us to decouple all subproblems, similar to \cite{Marthetal_JRSInterface_2015,Marthetal_JFM_2016}. We further conduct a shared memory OPENMP parallelization to solve the phase field equations and the orientation field equations via a parallel splitting method. Each linear system of equations is solved using the direct solver UMFPACK. Since the computational mesh has to be fine along the interface, adaptive mesh refinement is heavily used. However, using a single mesh for all variables is not appropriate in this case as e.g. the phase field variable $\phi_i$ only requires a fine resolution close to the zero level set of $\phi_i$ but not at the zero level sets of $\phi_j$ with $i \neq j$. The efficiency would go down if the number of cells increases if a single mesh would be used. The multi-mesh strategy, considered in \cite{Voigtetal_JCS_2012} for two meshes, overcomes these numerical problems and assigns a mesh to each phase field variable, which can be independently refined. In \cite{Lingetal_CMAM_2016,Marthetal_JFM_2016} the approach is extended to arbitrary meshes and validated for related problems.

\section{Simulations and results}\label{sec:simulations}
\subsection{Binary collisions of cells}
\begin{figure}[h]
\begin{center}
\includegraphics[width=0.78\textwidth]{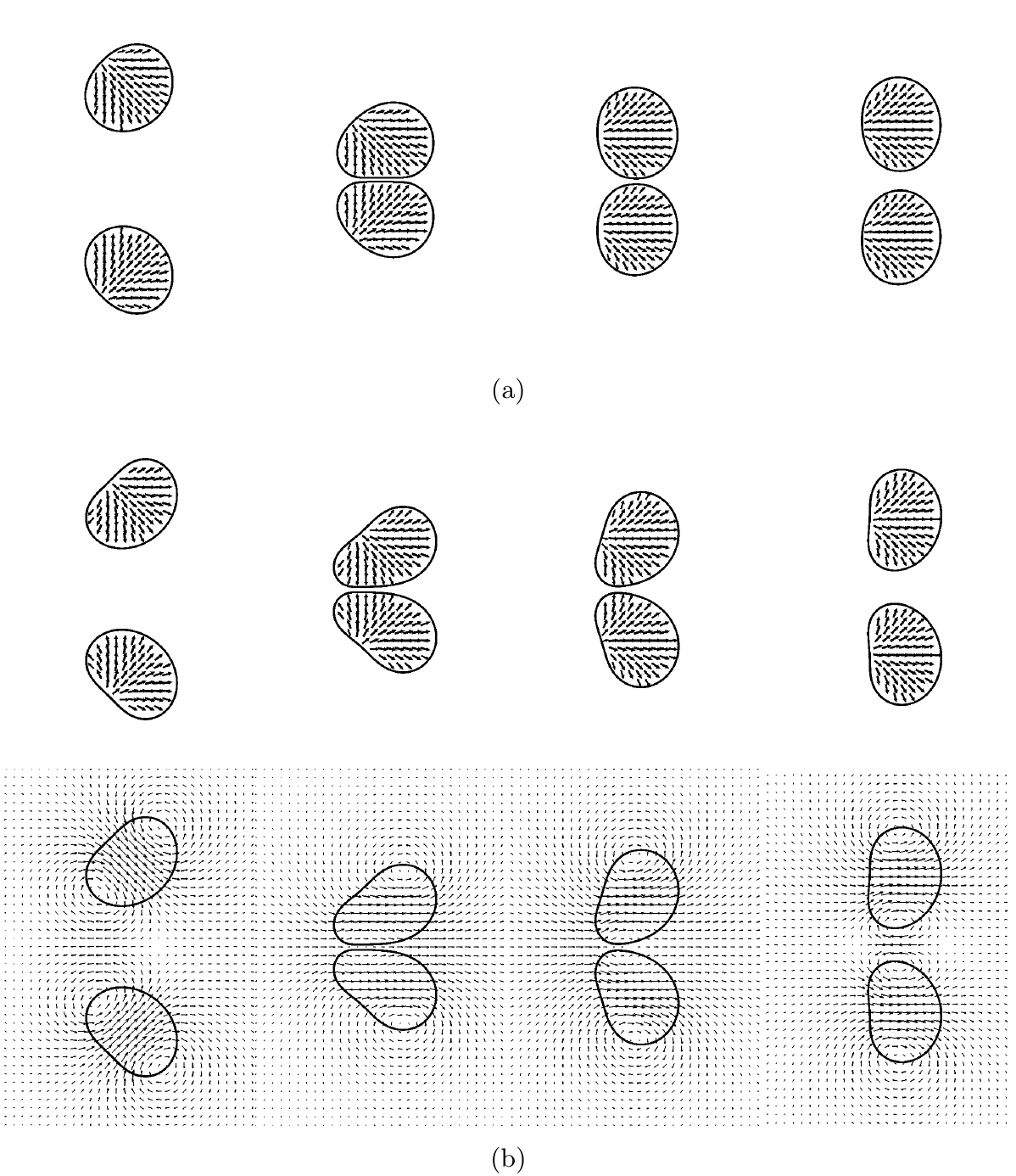}
\end{center}
\caption[Snapshots of a binary collision]{(a) non-hydrodynamic model. Shown are the cell shapes and the orientation fields. The parameters used are Ca$= 0.0281$, Be$= 0$, Pa$= 0.1$, In$=0.1125$, $c = 10$, $v_0 = 2.25$, $\beta = 0.5$, $\gamma = 1$, $\epsilon = 0.2$, $\kappa = 1$. (b) hydrodynamic model. Shown are the cell shapes and the orientation fields, together with the flow field. The parameters used are Ca$= 0.025$, Be$= 0$, Pa$= 0.1$, In$=0.1$, Fa$= 1$, Re$= 0.001$, $c = 10$, $v_0 = 3$, $\beta = 0.5$, $\gamma = 0.003$, $\epsilon = 0.2$, $\kappa = 1$, $\theta = 1$, $\xi = 0$. The time instances for both cases are $t = 3$, $17$, $30$ and $45$.}
\label{fig2}
\end{figure}

\begin{figure}[h]
\begin{center}
\includegraphics[width=0.8\textwidth]{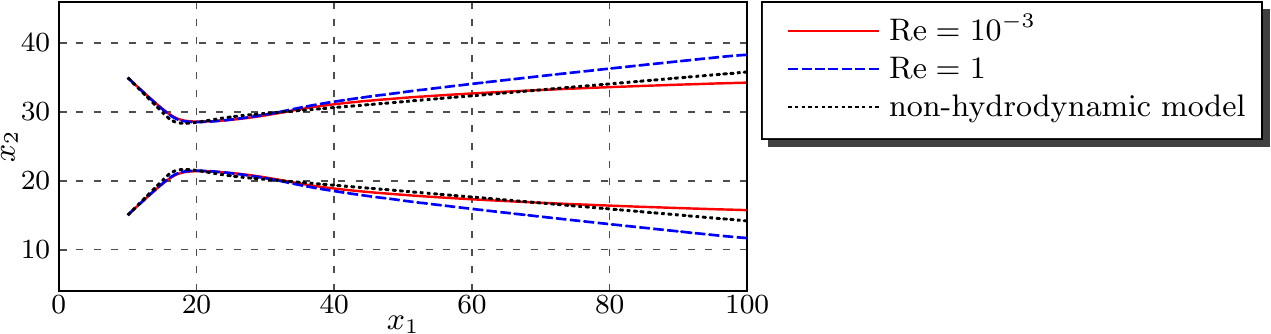}
\end{center}
\caption[Binary collision]{Center of mass trajectories for binary collision for the cases considered in Fig. \ref{fig2} and Re$ = 1$. (Online version in colour.)}
\label{fig3}
\end{figure}

We first study binary collisions of cells within a symmetric setup with a fixed incidence angle of $45^\circ$. Fig. \ref{fig2} shows snapshots of the cell shapes and orientation fields together with the flow field if appropriate. The cells deform at collision, the deformation influences the orientation fields which set the new directions for cell motion. For the hydrodynamic model each cell is accomplished with two vortices. Open collision the two vortices which come close to each other annihilate. This leads to a rotation of the cells and together with the deformation and reorientation of the orientation fields set the new directions for cell motion. In both cases, the non-hydrodynamic and the hydrodynamics case the coupling between the involved fields leads to partly inelastic collisions and alignment. However, the strength of the alignment strongly depends on various parameters. Fig. \ref{fig3} shows the center of mass trajectories for the non-hydrodynamic model and for the hydrodynamic model for different Reynolds numbers Re. The results show a tendency from more inelastic towards more elastic collisions for increasing Re. 

All simulations are performed within a two-dimensional computational domain of size $[0,50]^2$. Each cell has a size, corresponding to a circle with radius $R = 4$. We apply periodic boundary conditions in each direction. A systematic study of the influence of various parameters on alignment (not shown) reveals mainly the same qualitative dependencies for the hydrodynamic and the non-hydrodynamic model, even if the mechanism behind alignment significantly differs. The alignment is more efficient at small incidence angles and it is stronger for higher Capillary numbers Ca and smaller Polarity number Pa. Only the strength of the self-propulsion $v_0$ seems to have the opposite effect. While a larger value for $v_0$ leads to more elastic collisions in the non-hydrodynamic model, it leads to more in-elastic behavior in the hydrodynamic model. However, the effect is small if compared with the influence of the other parameters. The influence of the Bending capillary number Be is negligible. All other parameters are kept fixed. Clearly, the binary interaction behavior is beyond simple particle-based models, even if elastic deformations and/or hydrodynamic interactions are considered. The strength of alignment in the considered models is a result of the complex interplay between the cell shapes, viscosity, passive and active stresses, as well as actin polarizations and adhesion. The results further indicate the effect of the hydrodynamic interactions, with a tendency towards more elastic collisions for increasing Reynolds number Re. 

\subsection{Collective motion}\label{sec:collectmo}
\begin{figure}
\centering
\includegraphics[width=0.99\textwidth]{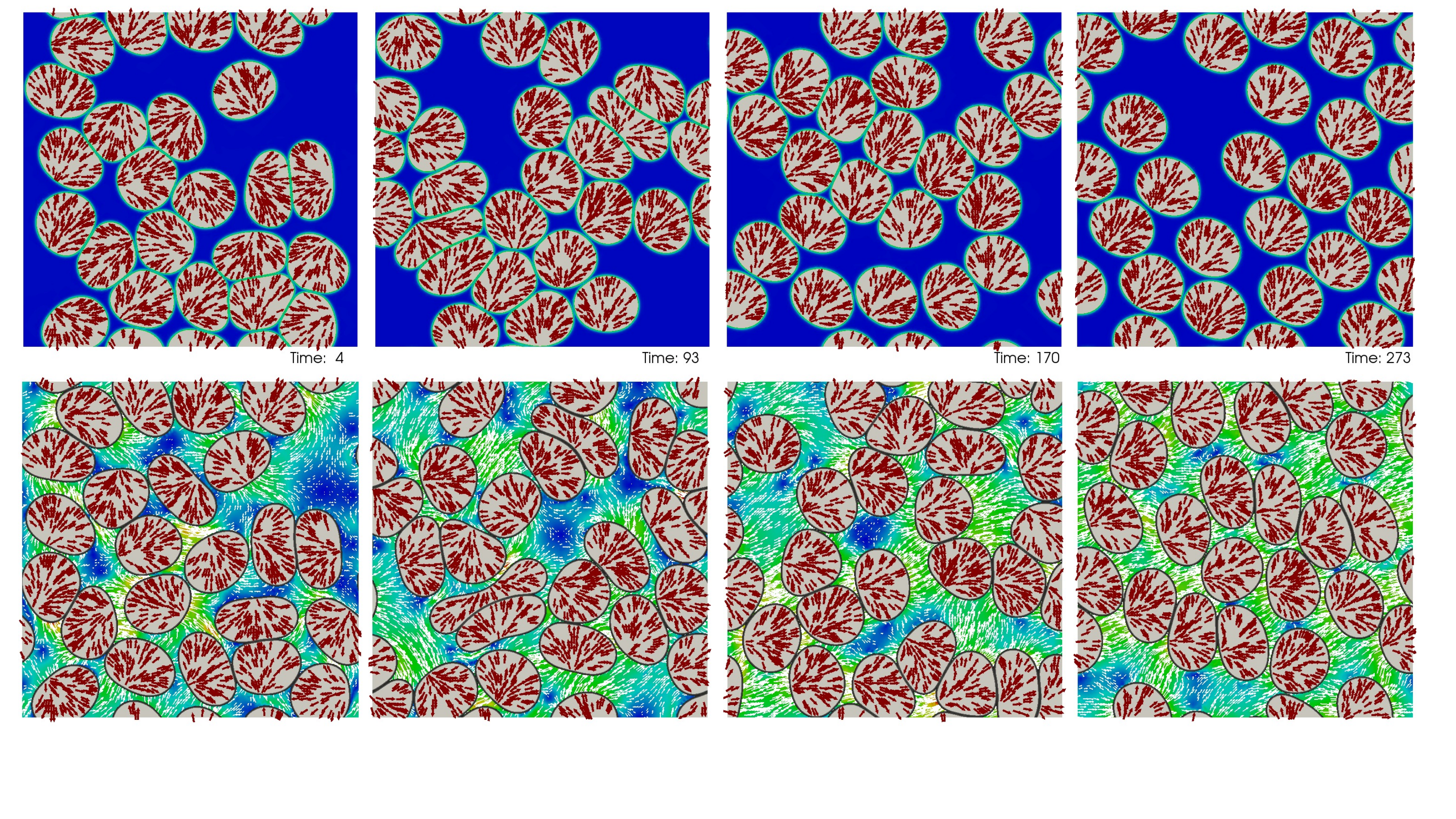}
\caption[Collective motion]{Snapshots of the cell shapes, orientation fields and fluid velocity, if appropriate. (top row) non-hydrodynamic model, (bottom row) hydrodynamic model. The snapshots correspond to the same times, shown in non-dimensional units. The parameters are the same as in Fig. \ref{fig2}. See also supplementary movie 1 and 2. (Online version in colour.)}
\label{fig4}
\end{figure}

We now investigate collective motion. For low cell densities collective motion is dominated by binary collisions. So from the previous results we might guess the onset of collective motion also within the hydrodynamic model, at least for low Reynolds numbers Re. To quantify the effect we introduce an order parameter 
\begin{align*}
\omega(t)=\frac{1}{N}|\sum_{i=1}^N  \frac{\mathbf{v}_i(t)}{|\mathbf{v}_i(t)|}|,
\end{align*}
with $\mathbf{v}_i$ the velocity vector of the $i$-th cell. The parameter $\omega$ is $1$ if all cells move in the same direction and $0$ if no correlation of the directions exists. Fig. \ref{fig4} shows snapshots of the evolution for $23$ identical cells, which initially move in random directions. The cell size now corresponds to a circle with radius $R = 4.5$. The domain sizes as well as all other parameters are as in the previous section with Reynolds number Re$= 0.001$.

\begin{figure}
\centering
\includegraphics[width=0.89\textwidth]{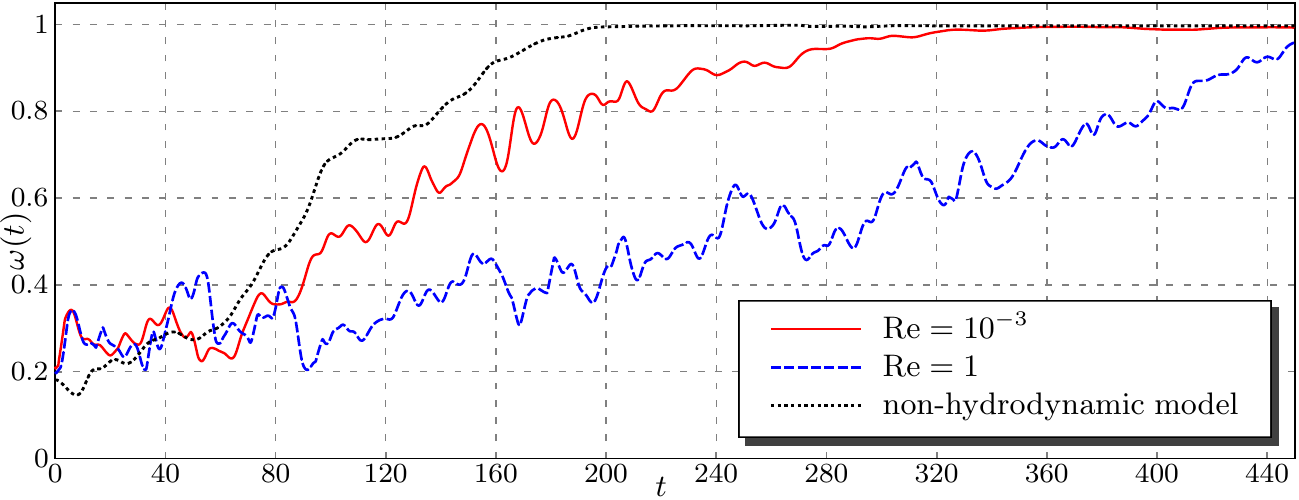}
\caption[Evolution of the order parameter]{The diagram shows the temporal evolution of $\omega$ for the non-hydrodynamic and the hydrodynamic model for two different Reynolds numbers Re. (Online version in colour.)}
\label{fig5}
\end{figure}

The result is quantified in Fig.~\ref{fig5}, which shows the evolution of $\omega$ for the non-hydrodynamic model and the hydrodynamic model for two different Reynolds numbers Re. 
These results for the non-hydrodynamic model confirm the findings in \cite{Loeberetal_SR_2015}: Without hydrodynamic interactions collision of deformable cells can lead to collective migration if the collisions are inelastic. This is even true if for each cell a separate orientation field is used and thus any diffusion/elastic interaction between these fields is impossible. The situation with hydrodynamics has not been analyzed before. The results indicate that also for low Reynolds numbers Re$ = 0.001$, which essentially corresponds to the Stokes regime and is the most relevant situation for substrate-based cell motility, collective migration can be observed. The time to reach collective motion is longer, but all simulations within this regime lead to persistent translational collective migration. Even if the mechanism is different, the analogy between inelastic binary collisions and collective migration seems to hold also for the hydrodynamic model with low Re. For Re$ =1$ the situation changes. The binary collision was more elastic and thus does not suggest collective migration. However, the more elastic collisions can not suppress collective migration only the time to reach this state is significantly increased. 

Increasing the viscosity of the cells $\eta(\text{cell})$ relative to the viscosity of the surrounding fluid $\eta$ (results not shown) has qualitatively no influence on these results. In both cases Re$ =0.0001$ and Re$ =1$ and $\eta / \eta(\text{cell}) = 0.1$ collective migrations is reached faster as for $\eta / \eta(\text{cell}) = 1$ and the fluctuations in $\omega(t)$ before reaching collective motion are reduced. 

These simulations indicate collective migration for deformable cells even under the influence of hydrodynamic interactions. In the low Reynolds number regime all performed simulations result in collective migrations. The effect seems to be as stable as without hydrodynamic interactions. Only for Re$ = 1$ the time to reach collective migration is significantly increased and even larger Re might be able to suppress the formation of collective motion.

\section{Conclusion}

We have developed a computational model for the collective migration of cells. On a single cell level, the model is based on the well-established mechanisms of cell motility accounting for actin polymerization, motor-induced contractility, and substrate adhesion. The model uses the hydrodynamic active polar gel theory \cite{Kruseetal_PRL_2000,Kruseetal_PRL_2004,Kruseetal_EPJE_2005} and is comparable to the approaches in \cite{Tjhungetal_PNAS_2012,
Ziebertetal_JRSInterface_2012,Giomietal_PRL_2014,Marthetal_JRSInterface_2015}. Each cell is treated individually using one phase field variable per cell. Cell-cell interaction is considered through an additional potential with a short range repulsive force as used and validated in \cite{Marthetal_JFM_2016,Lingetal_CMAM_2016}. The overall model only uses physical mechanisms, which do not need continuous regulatory effort. It describes details of the motility mechanism which allows to study the influence of many parameters on the dynamic behavior. The related non-hydrodynamic model \cite{Loeberetal_SR_2015} could already reproduces many experimentally observed phenomena. The overall question to answer is, if these phenomena persist under the influence of hydrodynamic interactions, which is controversially discussed \cite{Matas-Navarroetal_PRE_2014,Matas-Navarroetal_SM_2015,Tiribocchietal_PRL_2015}. On the level of detail, which is considered in this paper, the effect of hydrodynamic interactions has not been studied before. Our results on the collision of two cells lead qualitatively to the same results as in the non-hydrodynamic model \cite{Loeberetal_SR_2015}. These binary cell interactions may be quantified in terms of inelastic or elastic collisions. In the hydrodynamic model the variation of various parameters show the same tendency to one or the other as in the non-hydrodynamic case. However, with a stronger deformation of the cells and a more elastic behavior if the Reynolds number Re increases. As inelastic collisions has been reported as one indicator for collective migration \cite{Loeberetal_SR_2015}, these results suggest the onset of collective migration also if hydrodynamic interactions are taken into account, at least for low Re. The simulations with $23$ cells confirm this. All considered cases lead to persistent translational collective migration. Only the time to reach it differs and increases significantly with increasing Re. The considered parameters are Re$= 0.001$ and Re$=1$. Even larger Re, which might be able to suppress collective migration, are irrelevant for typical situation of substrate-based cell motility. These results provide valuable insight into the physics behind the biological processes in collective cell migration. It answers fundamental questions on collective motion for self-propelled particles and suggests some experimentally testable predictions. Can collective migration be found without cell-cell adhesion, is the effect stronger for cells with smaller membrane tension and larger elastic properties, as all predicted by our simulations, and can the effect of viscosity on collective migration be observed?

\vspace*{1cm}
WM and AV acknowledge support from the German Science Foundation through Vo899/11. We further acknowledge computing resources at JSC through grant HDR06. We also would like to thank the Isaac Newton Institute for Mathematical Sciences for its hospitality during the program "Coupling Geometric PDEs with Physics for Cell Morphology, Motility and Pattern Formation" supported by EPSRC Grant Number EP/K032208/1

\bibliographystyle{vancouver}      
\bibliography{sum}   

\end{document}